\documentclass[9pt,twocolumn,twoside]{pnas-new}
\usepackage{epstopdf}
\usepackage[capitalise]{cleveref}
\usepackage{textcomp}
\usepackage{soul}
\usepackage[none]{hyphenat}

\templatetype{pnasresearcharticle} 

\title{An \textit{in vitro} living system for streaming flow rectification}

\author[a,1]{Zhi Dou}
\author[a,1]{Liu Hong}
\author[b,c,1]{Zhengwei Li}
\author[a,1]{Fan Kiat Chan}
\author[a]{Yashraj Bhosale}
\author[a]{Onur Aydin}
\author[a]{Gabriel Juarez}
\author[a,g]{M. Taher A. Saif}
\author[a,d,e,f]{Leonardo P. Chamorro}
\author[a,g,h,2]{Mattia Gazzola}
\affil[a]{Mechanical Science and Engineering, University of Illinois at Urbana-Champaign, Urbana, IL 61801, USA}
\affil[b]{Biomedical Engineering, University of Houston, TX 77204, USA}
\affil[c]{Biomedical Science, University of Houston, TX 77204, USA}
\affil[d]{Aerospace Engineering, University of Illinois at Urbana-Champaign, Urbana, IL 61801, USA}
\affil[e]{Civil and Environmental Engineering, University of Illinois at Urbana-Champaign, Urbana, IL 61801, USA}
\affil[f]{Geology, University of Illinois at Urbana-Champaign, Urbana, IL 61801, USA}
\affil[g]{Carl R. Woese Institute for Genomic Biology, University of Illinois at Urbana-Champaign, Urbana, IL 61801, USA}
\affil[h]{National Center for Supercomputing Applications, University of Illinois at Urbana-Champaign, Urbana, IL 61801, USA}

\leadauthor{Dou}

\significancestatement{When a liquid vibrates around a small feature, inertial rectification generates precisely controllable steady flows. This phenomenon, known as viscous streaming, is attained via high-frequency external drivers and is employed in microfluidics for flow/particle/chemical manipulation. However, streaming may also be encountered at lower frequencies in biological settings of larger scale. Aquatic larvae or bacteria might then employ oscillating cilia or appendages to endogenously rectify surrounding flows, for feeding or locomotion, with no external assistance. Here, a living biological streamer is engineered to verify this possibility. Powered by muscle cells, our biostreamer rectifies surrounding flows, spontaneously or upon optical stimulation, illustrating both autonomy and control. This work thus synergistically integrates bio-hybrid technology and fundamental fluid mechanics, illustrating a complementary route to studying biologically actuated flows.}

\authorcontributions{Z.D., Z.L, and F.K.C. designed and perform research; L.H. and F.K.C. analyzed data; Z.D., F.K.C., G.J., and M.G. wrote the paper.}
\authordeclaration{The authors declare no competing interests.}
\equalauthors{\textsuperscript{1} Z.D., H.L, Z.L., F.K.C. contributed equally to this work.}
\correspondingauthor{\textsuperscript{2}To whom correspondence should be addressed. E-mail: mgazzola@illinois.edu}


\begin{abstract}
Small -- but finite -- fluid inertia can be leveraged to generate steady flows out of liquid vibrations around an immersed interface. In engineering, external high-frequency drivers (10$^2$--10$^5$ Hz) allow this inertial rectification phenomenon, known as viscous streaming, to be employed in micron-scale devices for precise flow control, particle manipulation and spatially controlled chemistry. However, beyond artificial settings, streaming may also be accessed by larger-scale biological systems pertaining to lower frequencies. Then, millimeter-size bacteria or larvae that oscillate cilia and appendages in the 1--10Hz range may be able to \textit{endogenously} rectify surrounding flows, for feeding or locomotion, removing the need for external actuators, tethers or tubings. In support of this hypothesis, here we demonstrate an \textit{in vitro} living system able to produce streaming flows, endogenously, autonomously and unassisted. Computationally informed, our biological device generates oscillatory flows through the cyclic contractions of an engineered muscle tissue, shaped in the form of a ring and suspended in fluid within a Micro-Particle Image Velocimetry setup, for analysis. Flow patterns consistent with streaming simulations are observed for low-frequency muscle contractions (2--4Hz), either spontaneous or light-induced, illustrating system autonomy and controllability, respectively. Thus, this work provides experimental evidence of biologically powered streaming in untethered millimeter-scale living systems, showcasing the utility of bio-hybrid technology for fundamental and applied fluid mechanics.

\end{abstract}

\dates{This manuscript is compiled on \today}
\doi{\url{www.pnas.org/cgi/doi/10.1073/pnas.XXXXXXXXXX}}

\begin{document}

\maketitle
\thispagestyle{firststyle}
\ifthenelse{\boolean{shortarticle}}{\ifthenelse{\boolean{singlecolumn}}{\abscontentformatted}{\abscontent}}{}

\dropcap{W}hen a liquid vibrates around an immersed small feature, a solid inclusion or a bubble, fluid inertia and local concentration of stresses conspire to produce a steady flow out of the underlying unsteady field. This inertial rectification phenomenon, known as viscous streaming \cite{holtsmark1954boundary,lane1955acoustical, bertelsen1973nonlinear}, represents an efficient and robust flow machinery to manipulate fluid, suspended particles, and chemicals \cite{lutz2003microfluidics, lutz2006hydrodynamic, lutz2005microscopic, bhosale2021multi}. Depending on oscillation frequency, kinematic viscosity, body size and shape, a variety of streaming flow topologies can be generated \cite{bhosale2021multi, chan2021three, bhosale_parthasarathy_gazzola_2020, parthasarathy2019streaming}, conveniently partitioning the fluid domain into neatly separate recirculating regions.

Microfluidics has most prominently capitalized on viscous streaming for particle \cite{parthasarathy2019streaming,klotsa2007interaction,klotsa2009chain} and cell transport \cite{marmottant2004bubble}, sorting \cite{thameem2016particle}, separation \cite{wang2011size} or concentration \cite{ookawara2010process}, as well as for chemical mixing \cite{lutz2003microfluidics}. Due to the miniaturization focus and use of micron-scale features \cite{thameem2016particle,thameem2017fast}, these devices must employ high flow oscillation frequencies ($10^{2} - 10^{5}$ Hz) typically realized through an external driver, for example, a loudspeaker or a piezo stack \cite{vishwanathan2019steady, vishwanathan2021}. Micro-scale artificial settings and high-frequency actuation have thus come to define viscous streaming.

However, viscous streaming may also be accessed by biological systems of larger scale operating at lower frequencies. Then, a multitude of millimeter-scale aquatic creatures that oscillate cilia or appendages in the $\sim1-10$ Hz range may be able to autonomously rectify surrounding flows, for feeding or locomotion. Bacteria \cite{spelman2017arbitrary} and larvae \cite{gilpin2020multiscale} have been indeed linked to flow patterns and velocities ($\sim 10^1-10^2\mu$m/s) potentially consistent with streaming. Nonetheless, because of the small velocities involved, the establishment of steady streaming flows and their rigorous streaming identification requires acquiring and averaging flow data for (up to) tens of seconds, which renders challenging or altogether precludes the use of freely swimming samples. While these may be fixed in place between glass slides \cite{gilpin2017vortex}, resulting flows are artificially two-dimensional, weakening any firm conclusion relative to natural three-dimensional settings. Thus, the ability of millimeter-scale creatures to actively produce streaming in their habitat remains to be experimentally verified.

\begin{figure*}[t]
\centering
\includegraphics[width=\textwidth]{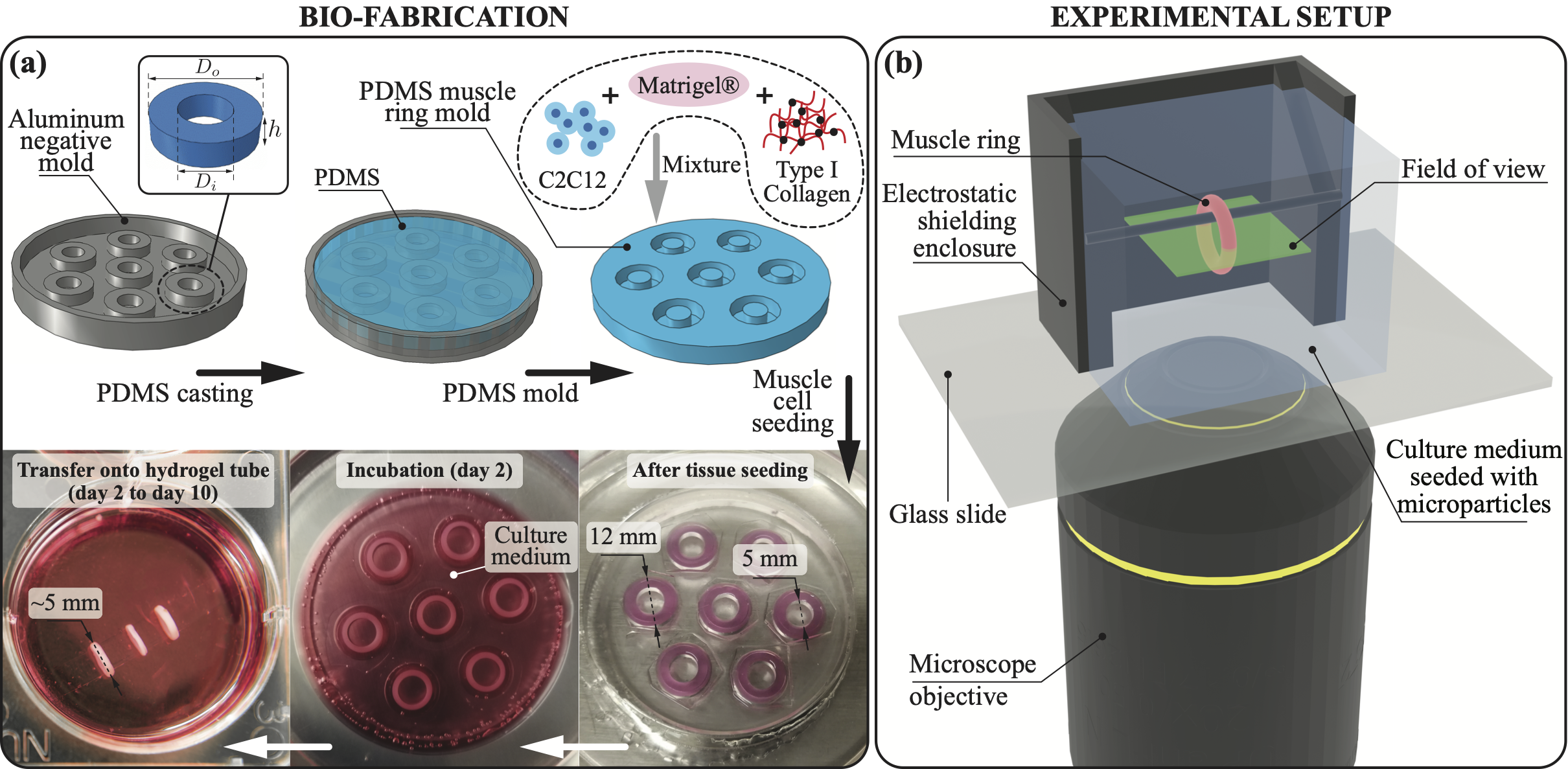}
\caption{Bio-fabrication of muscle rings and streaming experimental setup. (a)  An aluminum negative mold with torus extrusions of inner diameter $D_i=5$~mm, outer diameter $D_o=12$~mm and depth $h=3$~mm is machined. These extrusions define the dimensions of the wells in the PDMS mold. PDMS precursor is poured into the mold and allowed to polymerize at 60 $^\circ$C. After sterilizing and functionalizing (Methods) the surface of the detached PDMS mold, a mixture containing muscle cells, Matrigel ECM, and type-I collagen is injected into the mold, and culture media is added after the cell-gel mixture has polymerized at 37 $^\circ$C. When the rings have compacted for two days, they are transferred onto hydrogel tubes for differentiation until experiment. (b) A machined aluminum enclosure that is open on both top and bottom is glued onto a microscope glass slide and filled with cell culture media mixed with polystyrene tracer particles of diameter 6--8~$\mu$m and specific gravity of 1.05. The inner dimension of the enclosure is 1~cm$\times$1~cm$\times$1~cm and two of its opposing walls present cutouts at their centers, to hold a thin aluminum rod (1~mm diameter). For experiments, a muscle ring is transferred from the hydrogel tube onto the rod and suspended in the middle of the box. The entire apparatus is secured on the stage of a microscope integrated with a cell culture incubator and attached to a high-speed camera focused on the midplane of the ring for recording. For optical stimulation, the tip of an optical fiber connected to a laser diode is placed above the enclosure to produce light pulses.}
\label{fig:1}
\end{figure*}

Here, we explore this possibility via a living biomimicry, engineered to circumvent the above difficulties, realized \textit{in vitro} out of muscle cells (\cref{fig:1}a), and integrated within a Micro-Particle Image Velocimetry ($\mu$PIV) system (\cref{fig:1}b). The resulting bio-hybrid platform proves to be convenient for our inquiry. Indeed, muscle tissue contracts spontaneously at a natural frequency, which leads to a simple design where oscillations are self-powered, removing the need for external drivers, wires or tubings. The possibility of rendering cells light-sensitive \cite{deisseroth2011optogenetics} further provides the option of non-invasively controlling actuation frequency via optical inputs \cite{raman2016optogenetic,raman2017modular}, ridding the flow field of disturbances caused by otherwise electric stimulation. Additionally, by pouring cells and extracellular matrix (ECM) in a mold (\cref{fig:1}a), engineered muscles of different geometry can be formed \cite{pagan2019engineering}. We then realize our freely contracting device in toroidal shape and suspend it in bulk liquid by simply hanging it to a thin submerged horizontal rod (\cref{fig:1}b), altogether removing the need for fixing between glass slides, and thus preserving flow three-dimensionality. 


Computationally informed, a ring-shaped biological streamer made of skeletal muscle tissue is then shown to indeed generate streaming flows. The streamer works autonomously through its spontaneous contraction, or is remotely controlled via optical stimulation, demonstrating untethered, biological, millimeter-scale flow rectification.



\section*{Setup}

\subsection*{Fundamentals of viscous streaming}
Viscous streaming \cite{holtsmark1954boundary,bertelsen1973nonlinear}, an inertial phenomenon, arises when a small feature (typically a solid inclusion or a microbubble) of characteristic length $a$, is immersed in a liquid of kinematic viscosity $\nu$, that oscillates with angular frequency $\omega$ and amplitude $A\ll a$. Inertial rectification, facilitated by stress concentration at the boundaries of the immersed body, results in steady flows that can be effectively leveraged to manipulate fluid, suspended particles, and chemicals \cite{lutz2003microfluidics, lutz2006hydrodynamic, lutz2005microscopic, parthasarathy2019streaming, bhosale2021multi}. For simple geometries of uniform curvature (cylinders, spheres), viscous streaming flows have been theoretically, computationally, and experimentally well characterized as function of the Stokes layer thickness $\delta_\text{AC}\sim\mathcal{O}(\sqrt{\nu/\omega})$ and length-scale $a$ \cite{holtsmark1954boundary,riley2001steady,lutz2005microscopic,lane1955acoustical,bhosale_parthasarathy_gazzola_2020}. More recently, the use of multi-curvature streaming bodies has revealed rich flow repertoires, furthering opportunities in transport, separation, and assembly \cite{parthasarathy2019streaming,bhosale_parthasarathy_gazzola_2020,chan2021three,bhosale2021multi,bhosale2022soft}.

\subsection*{Biostreamer fabrication and flow imaging integration}

We aim to build a millimeter-scale, 3D muscle tissue to harvest contractile motions for the generation of oscillatory flows. All the while, the geometry of the tissue itself is selected and leveraged to modulate identifiable streaming fields, facilitate biofabrication, $\mu$PIV imaging, and flow analysis.

Here, skeletal muscle myoblasts are chosen for their ability to form 3D tissue and fuse into muscle fibers (myotubes) that align spontaneously along stress fields, thus providing directed contractions \cite{neal2014formation}. This can be achieved by allowing the tissue to anchor on a pair of posts \cite{aydin2019neuromuscular,aydin2020development}, providing axial tension thus longitudinal alignment and contraction, or by letting the tissue wrap around a single cylindrical pillar \cite{uzel2014microfabrication}, generating tangential stresses and tangentially aligned fibers. These, upon contraction, cause the circumference of the tissue to shorten, resulting in overall radial actuation \cite{li2019biohybrid,li2022adaptive}.

The latter approach is particularly suited to our investigation (\cref{fig:1}a). Indeed, the cylindrical pillar minimizes uneven stress distributions, favoring regular fiber organization, tissue shape, and uniform radial contractions. This, in turn, is conducive of reproducible flow responses. Further, geometric axisymmetry implies the existence of a midplane where the flow becomes two-dimensional, simplifying flow imaging and analysis. Finally, a ring shape can be conveniently suspended in bulk liquid by simply sliding it onto a thin submerged rod (\cref{fig:1}b).

To fabricate our muscle ring (\cref{fig:1}a), a polydimethylsiloxane (PDMS) mold is first created by pouring liquid precursor into a machined negative mold and allowing it to polymerize. A mixture of C2C12 mouse skeletal muscle myoblasts, ECM, and type-I collagen is then cast into the PDMS mold, where the myoblast-laden gel compacts into rings \cite{bell1979production}. They are subsequently transferred onto a hydrogel tube submerged in culture medium, to induce differentiation, during which contractile myotubes form and align. Because of tube compliance and muscle internal tension, the diameter of the torus shrinks of typically 1--2~mm, before stabilizing.

\begin{figure*}[ht!]
\centering
\includegraphics[width=\textwidth]{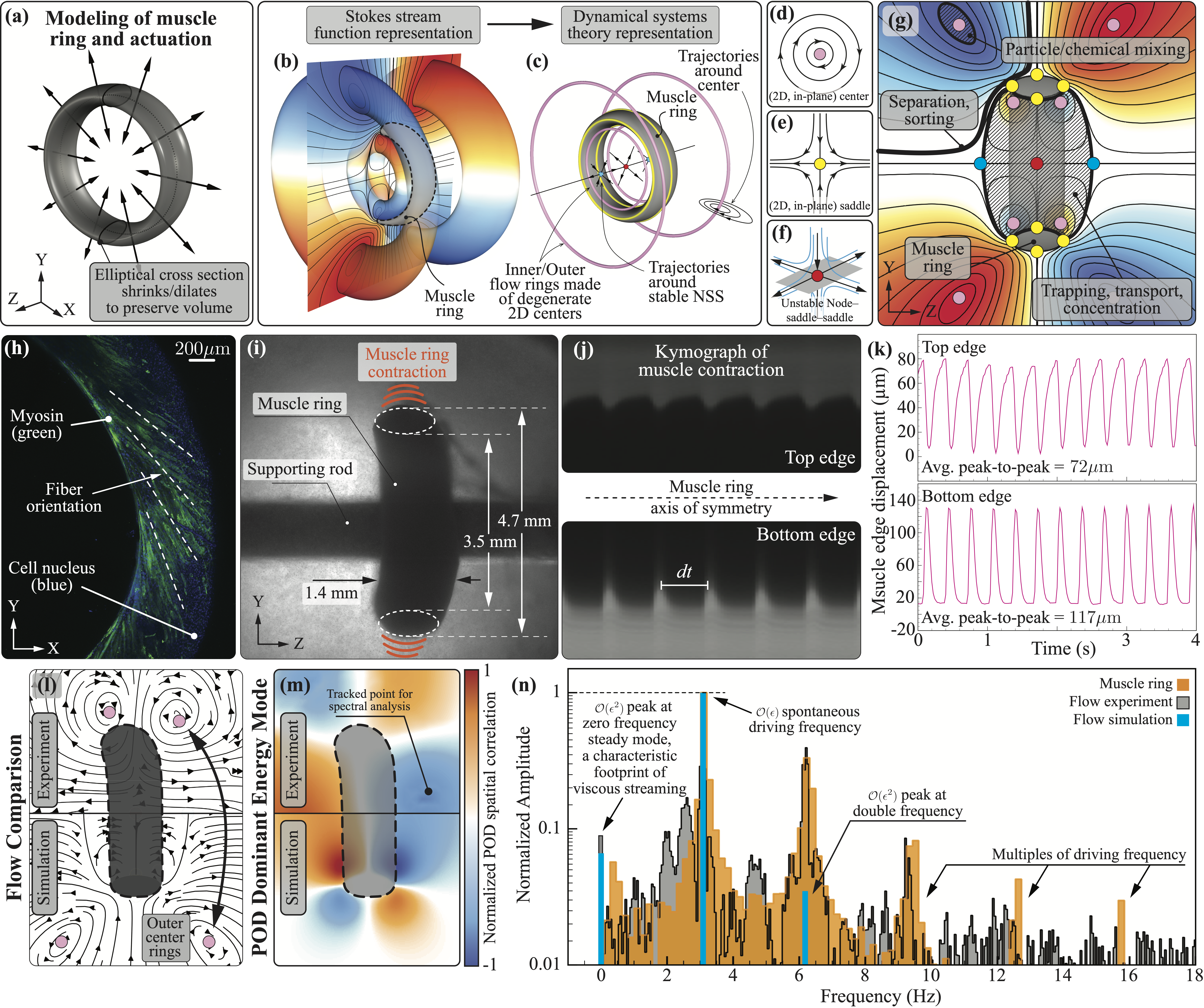}
\caption{Computational design of the biostreamer and autonomous streaming. 
(a) Modeling of the muscle ring and its radial contractions. The computational muscle ring and surrounding liquid environment are characterized, consistent with experiments, by the non-dimensional Stokes layer thickness $\delta_{AC} / a = 0.11$, where the length scale $a$ is the depth of the muscle ring.
(b-c) Numerical simulation of the streaming flow around the muscle ring. The visualization of the flow field is presented using two different methods: (b) Stokes stream function  and (c) dynamical system representation.
(d-f) Observed critical points from simulation and illustrations of their corresponding local flow patterns.
(g) Expected flow structure at the midplane of the muscle ring from simulation. Critical points are marked by circles with colors corresponding to (d-f). Colour contours on the presented plane indicate regions of clockwise (blue) and counter-clockwise (orange) recirculating fluid. Regions near the critical points where potential particle manipulations can be achieved are encircled in bold lines. 
(h) Confocal fluorescent $z$-stack image of a section of a muscle ring showing the circumferentially aligned myotubes (green) and cell nuclei (blue). 
(i) Bright-field image showing the muscle ring hanging on the thin rod. The ring has measured inner diameter of 3.5~mm, outer diameter 4.7~mm, and depth 1.4~mm. Orange curves illustrate the radial muscle contractions.
(j) Kymograph of the top and bottom edge of the muscle ring showing its periodic contractions over 2 seconds.
(k) Plots of the top and bottom edge displacement due to muscle actuation over 4 seconds.
(l) Time-averaged streamlines of the upper half of the muscle ring from experiment juxtaposed with streamlines of the lower half of the muscle ring from numerical simulation, showing the matching centers and recirculating regions. 
(m) Dominant first POD mode extracted using $x$ velocity components ($u$) from experiments (upper half) juxtaposed with those from simulations (lower half), demonstrating coherent regions from comparatively high correlation zones (flow locations of the same color indicate that local fluid particles move coherently in a correlated fashion; orange/blue colors indicate positive and negative values; the change in sign indicates an opposite sense of motion). 
The location used for obtaining the experimental flow spectrum is marked. 
Similar results are observed when analyzing at other flow locations (Fig. S3b) and when considering the $y$ velocity component $v$ (Fig. S3d). 
(n) Frequency spectrum of the experimental flow ($u$), simulated flow, and muscle actuation. The normalized amplitude is presented on a logarithmic scale. 
}
\label{fig:2}
\end{figure*}

For flow measurement and analysis, a $\mu$PIV system is realized (\cref{fig:1}b). A rectangular aluminum enclosure is machine fabricated and bonded on a glass slide to hold culture medium laden with tracer particles of diameter 6--8~$\mu$m. Besides serving as a container, the aluminum box also shields the tracers from external electric fields, minimizing drift caused by their slight negative charge. Two of the box opposing walls present cutouts at their centers to hold a thin aluminum rod onto which the muscle ring is slid. The box has inner dimensions of 1~cm$\times$1~cm$\times$1~cm. This choice is dictated by the need to minimize convection, which is observed to be significant (compared to expected streaming velocities) for larger volumes. A direct consequence of using a relatively small box is that flow boundary effects come into play. Nonetheless, we note that streaming is robust to moderate confinement, as demonstrated in periodic lattices both experimentally and numerically \cite{bhosale2021multi}. We thus expect that an eventual streaming field generated by the muscle ring will be topologically unaffected by the presence of the walls, although we do anticipate a geometric compression of the main flow structures towards the center of the enclosure. This, in turn, may facilitate flow visualization. Indeed, by bringing characteristic flow features closer together, the area to be imaged is reduced, allowing improved resolution and $\mu$PIV reconstruction accuracy.


The box and glass slide are mounted on the $x$-$y$ translation stage of an Olympus IX81 inverted microscope equipped with a $2\times$ objective. The differentiated biostreamer is then transferred into the $\mu$PIV system, and we adjust the stage such that the muscle ring is at the center of the field of view. We focus the microscope to the midplane of the toroidal tissue. A Hamamatsu digital camera connected to the microscope is utilized to capture images at 50 Hz, which are paired with a 5-frame step size and integrated with spatial cross correlation to produce velocity field data \cite{hart1998elimination,keane1992theory}.
The biostreamer is thus allowed to contract spontaneously, or upon stimulation by an external optical fiber setup, generating an oscillatory flow in its surrounding. In the case of optical stimulation, tissue biofabrication entails the use of C2C12 cells with a mutated variant of the blue light-sensitive ion channel, Channelrhodopsin-2 (ChR2). Myotubes differentiated from these myoblasts selectively respond to precise light wavelengths, in our case 470~nm \cite{raman2016optogenetic}. To reach the illumination intensity required for muscle contraction \cite{sakar2012formation,bryson2014optical,klapoetke2014independent}, we connect the fiber to a laser diode source producing light pulses of intensity 1.6~mW/mm$^2$. The specific combination of light wavelength and intensity prevents the $\mu$PIV illumination system from interfering with the biostreamer activity.

\section*{Results}

\subsection*{Computational design of the biostreamer} Streaming literature has almost exclusively examined the use of bodies without holes (genus-0), unlike our muscle ring (genus-1). While in one recent instance \cite{chan2021three} a rigid torus was computationally investigated for externally driven flow oscillations (perpendicular to the body axis of symmetry), radial self-contractions have not been considered so far. Further complicating the design specifics of our system, 3D skeletal muscle constructs have been demonstrated in a range of dimensions and actuation conditions. Viable muscle rings have been indeed reported for inner diameters ranging from 0.5~mm to 14~mm \cite{okano1997tissue,gwyther2011engineered,li2019biohybrid,li2022adaptive,pagan2018simulation}, with wall thicknesses between 0.5~mm and 4~mm \cite{gwyther2011engineered,pagan2018simulation}. Spontaneous contractions have been found to occur within the frequency range of 1~Hz to 3~Hz, and oscillation amplitudes observed to vary between 20~$\mu$m and 500~$\mu$m. \cite{li2019biohybrid,li2022adaptive,pagan2018simulation}. Additionally, optogenetic C2C12 muscle tissues have been shown to produce consistent force responses to optical stimulation across the 1~Hz to 4~Hz range \cite{raman2016optogenetic}.


To provide insight into the behavior of our coupled flow-muscle system, and in search of design solutions able to robustly generate streaming, we simulate different biostreamer instances within the above outlined, realizable parameter space. We model the muscle ring as a solid undergoing prescribed radial oscillations (while conserving volume, \cref{fig:2}a), immersed in an incompressible viscous fluid of unbounded domain. The coupled system is numerically discretized and simulated using a remeshed vortex method coupled with Brinkmann penalization (Methods), an approach \cite{gazzola2011simulations} demonstrated in a variety of settings \cite{bhosale2021remeshed,gazzola2011simulations,gazzola2012c,gazzola2012flow,gazzola2014reinforcement,gazzola2016learning}, including streaming \cite{parthasarathy2019streaming,bhosale_parthasarathy_gazzola_2020,chan2021three}.


In simulations, the investigated biostreamers are observed to produce qualitatively similar streaming flows, although quantitative details vary depending on muscle geometry and actuation properties. A representative instance is reported in \cref{fig:2}b-g. On the left (\cref{fig:2}b), isosurfaces of the time-averaged Stokes stream function (blue and orange representing clockwise and counter-clockwise flow rotations) highlight the presence of four recirculating annular regions (two on each side of the muscle ring). Two large regions are found to envelope the biostreamer, while a pair of smaller, nested ones approximately lie within the muscle itself. On the right (\cref{fig:2}c) is the corresponding dynamical representation, whereby critical points (characterized by zero velocity) are extracted from the flow and classified based on their stability properties (through the Jacobian of the local velocity field).



The flow is found to be organized around four types of critical points: centers (pink -- \cref{fig:2}d), saddles (yellow -- \cref{fig:2}e) and attractive (blue) / repulsive (red) node-saddle-saddle (NSS -- \cref{fig:2}f). In our axisymmetric case, both centers and saddles are 2D degenerate \cite{strogatz2018nonlinear}, since the surrounding local flow has no out-of-plane (i.e. azimuthal) component. These degenerate points, when mapped to 3D space, form four continuous center-rings (pink) and four continuous saddle-rings (yellow), as illustrated in \cref{fig:2}c. They collectively define the four annular recirculating regions described above.

The extracted flow skeleton underscores the structures that are most likely to be captured experimentally. A streamfunction slice through the midplane (\cref{fig:2}g) mimics experimental visualization conditions and provides intuition. The locations at which the rings cross the midplane are identified by dots of corresponding colors, elucidating the role of centers and saddles in shaping the flow while highlighting their practical utility. Specifically, centers initiate recirculation and can be employed to attract/retain particles \cite{chong2013inertial,chong2016transport} and mix fluids, while saddles (and connecting streamlines) partition the flow, enabling particle separation and sorting \cite{lutz2003microfluidics,thameem2016particle} or spatially controlled chemistry \cite{lutz2003microfluidics,lutz2006characterizing}. Based on \cref{fig:2}g, we anticipate that inner rings, being tucked within the muscle, will be difficult to image experimentally. Further, repulsive (red) and attractive (blue) saddles, that lie along the axis of symmetry in unperturbed simulations, may not be found on the experimentally imaged midplane, as a consequence of implementation imperfections and perturbations induced by the aluminum rod. Thus, the two external center-rings are likely the most robustly detectable structures. We then expect to observe, projected on the $\mu$PIV imaging plane, four centers, located around the muscle ring and associated with recirculatory flows.

Overall, we numerically find that muscle rings of approximately $\sim$4~mm inner diameter and $\sim$5~mm outer diameter may be ideal candidates for experimental realization. Indeed, at typical oscillation amplitudes ($\sim 100$~$\mu$m) and frequencies (1~Hz to 4~Hz), such rings are observed to generate streaming velocities ($\sim 10~\mu\text{m/s}$) significantly larger than background disturbances ($\sim 1~\mu\text{m/s}$). Further, key flow structures are predicted to closely surround the muscle, thus falling within a compact field of view, while remaining sufficiently separated for experimental detection.

\subsection*{Autonomous streaming}
We grow the biostreamer targeting the computationally identified specifications. A PDMS toroidal mold of 5~mm inner diameter, 12~mm outer diameter, and 3~mm depth is fabricated and the mixture of muscle cells and ECM is seeded (\cref{fig:1}a). Mold dimensions are empirically determined to account for compaction, differentiation, and associated shrinking, leading to muscles of the approximately desired geometry ($\sim$4~mm inner, $\sim$5~mm outer diameters). Confocal fluorescent imaging also confirms myotubes circumferential alignment, for radial contraction (\cref{fig:2}h).

Upon transfer to the $\mu$PIV system, the rings’ specific dimensions and spontaneous contraction amplitudes/frequencies are determined by tracking and averaging top and bottom edge displacements across 1,000 camera images acquired at 50~Hz. For the representative muscle of \cref{fig:2}h-n, we measure a 3.5~mm inner diameter, 4.7~mm outer diameter, 1.4~mm tissue depth, 3.1~Hz dominant contraction frequency, and 72~$\mu$m/117~$\mu$m average peak-to-peak top/bottom amplitudes. Obtained amplitudes and frequencies are reflected in the kymographs of \cref{fig:2}j, over 2 seconds, and in the time-varying edge displacements of \cref{fig:2}k, over 4 seconds.

To facilitate comparison, we input these muscle-specific parameters into our simulations and analyze both experimentally and computationally obtained flows. We first consider experimental steady streamlines, depicted in \cref{fig:2}l, top half. These are obtained by averaging recorded instantaneous velocities in the midplane, over 8 contraction cycles. This procedure removes (approximately) the flow's oscillatory components, revealing the underlying rectified, steady streaming field. As anticipated, we observe two distinct centers and corresponding flow recirculation regions. These features are also visible in the mirrored streaming field obtained from simulations, revealing close agreement (\cref{fig:2}l, bottom half). We note the slight asymmetry that characterizes experiments, on account of ring geometry and actuation imperfections as well as due to the presence of the aluminum rod.  These perturbations are also likely responsible for the missing NSS along the ring axis of symmetry, features that we previously identified as non-robustly observable. We further highlight how in experiments, centers are found closer to the biostreamer than in simulations.  This is consistent with the intuition that wall effects (enclosure) would result in flow geometric compression. Finally, we note that the overall streaming response is found to be repeatable, with associated features seen across cyclic muscle contractions and samples (Fig. S1, S3).

We proceed with identifying additional hallmarks of streaming, through a combination of Proper Orthogonal Decomposition (POD) \cite{sirovich1991analysis} and flow spectral analyses. POD considers space-time correlations in the velocity field and extracts energetic coherent structures lingering in the flow. We can employ this information to determine flow locations highly representative, in terms of their spectral fingerprint, of the underlying flow. We thus consider experimental and simulated midplane velocities over 20 seconds, extract the corresponding dominant POD mode (\cref{fig:2}m, 90\% of total energy, orange/blue intensity correlates to local flow energy), and sample it at different coordinates (\cref{fig:2}m and S3) at which spectral analysis of raw velocity data is performed.

In frequency domain, streaming flows present a well-known signature: a driving-frequency main peak capturing first order $\mathcal{O}(\epsilon)$ oscillatory effects, a zero-frequency peak (about ten times smaller) capturing second order $\mathcal{O}(\epsilon^2)$ streaming rectification, and a series of peaks of exponentially decaying magnitude at multiples of the driving frequency, corresponding to increasingly higher-order effects \cite{holtsmark1954boundary}. 
Our data are found to quantitatively recapitulate this structure, as illustrated in \cref{fig:2}n where we report the spectra of measured (black) and simulated (blue) velocities at the representative location marked as \textcircled{$\times$} in \cref{fig:2}m. 
Additionally, for reference, we plot the spectrum of the muscle top edge displacement (orange) of \cref{fig:2}k. As can be noticed, the measured/simulated flows indeed respond to the input muscle frequency (peak at 3.1~Hz) and produce rectification (peak at 0~Hz). The relative magnitude of these two peaks is also in line with theory, with rectification about one order of magnitude weaker, on account of being a second order effect.

We also note the presence of the characteristic peaks at multiples of the driving frequency. In simulations (blue bars), as expected from theory, the peak corresponding to the first frequency doubling ($\sim$6~Hz) presents a magnitude comparable to the rectification peak, consistent with the fact that this is a second order effect too. Remaining frequency-multiples peaks are also detected from simulated flows (Fig. S2), although they are not visible in \cref{fig:2}n given their negligible normalized amplitudes (that fall below the plotted lower bound), on account of their exponential decay \cite{holtsmark1954boundary}.

The experimental flow response (black bars) does present a similar structure, whereby driving, rectification and cascading peaks are all found at the theoretically and computationally expected frequencies. However, the relative magnitude of the peaks at frequencies multiples of the driving one, is significantly stronger than predicted. This is not necessarily surprising for the following complementary reasons. First, the muscle is soft and can deform in response to the flow, unlike in theory where the body is considered to be rigid or in simulations where the ring's deformations are imposed. Disregarding body compliance leads to underestimating the strength of the streaming response, as recently demonstrated in \cite{bhosale2022soft} whereby flow-induced elastic deformations are shown to be an additional source of streaming. Second, frequency multiples might originate from the tissue itself on account of its non-linear mechanical response to a driving oscillatory forcing (in this case cyclic contractions), through a mechanism mathematically similar to fluids \cite{holtsmark1954boundary}. This effect, pertaining to the solid phase, might also couple with the flow, strengthening both responses. That a combination of these factors may be at play is suggested by the muscle spectrum (orange bars) of \cref{fig:2}n, whereby deformations are found to precisely sync with the flow, presenting a set of highly structured harmonics that are unlikely to be the result of spontaneous, higher-frequency tissue contractions. While an explanation for the observed peak enhancement is provided, dissecting the exact mechanisms at play is beyond the scope of this work and not strictly necessary to confirm the presence of streaming or demonstrate the utility of our bio-hybrid platform.

Overall, given the observed qualitative and quantitative agreement between experiments, simulations and theory, we conclude that our tissue-engineered biostreamer is indeed capable of generating streaming autonomously.

\begin{figure*}[t!]
\centering
\includegraphics[width=\textwidth]{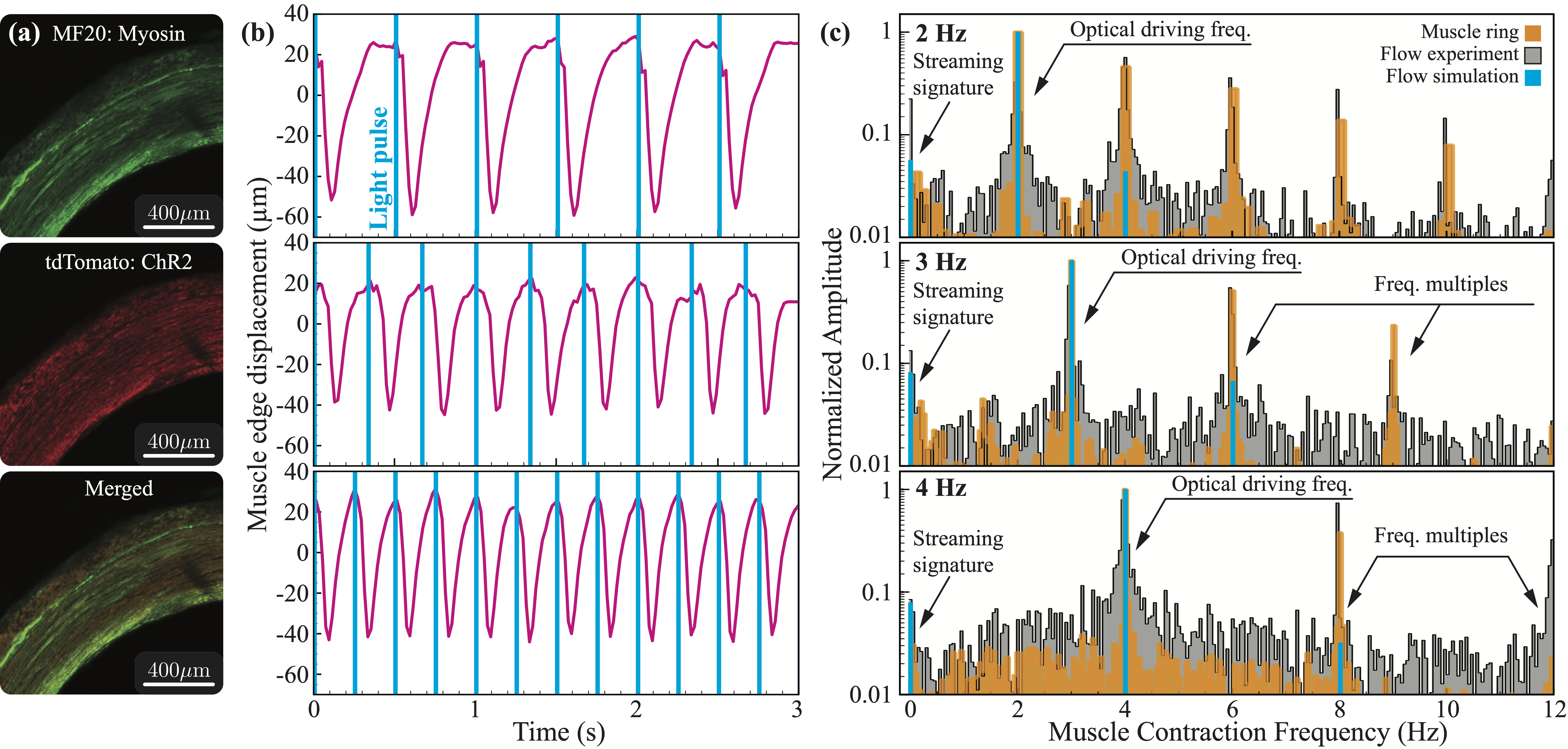}
\caption{Optically controlled streaming. 
(a) Confocal fluorescent $z$-stack images of a section of a muscle ring showing the myotubes (MF 20, green) and the ChR2 marker (tdTomato, red) at the same locations. 
(b) Muscle edge displacement of the top half of the muscle ring over 3 seconds (pink) and corresponding light stimulation pattern made of pulses of 10~ms duration (blue) at different frequencies. (top) 2~Hz stimulation results in an average peak-to-peak amplitude of 81~$\mu$m, (middle) 3~Hz stimulation and 59~$\mu$m peak-to-peak amplitude, and (bottom) 4~Hz stimulation and 61~$\mu$m peak-to-peak amplitude. (c) Frequency spectrum of the experimental flow ($u$), simulated flow, and muscle actuation at optical driving frequencies of (top) 2~Hz, (middle) 3~Hz, and (bottom) 4~Hz. The normalized amplitude is presented on a logarithmic scale. 
}
\label{fig:3}
\end{figure*}

\subsection*{Optically controlled streaming}
Now, we demonstrate controllability of the native function of the muscle ring by exploiting muscle contractions in response to an external light stimulus. 
To control the actuation frequency using optical stimulation, we fabricate our biostreamer following the same process of the autonomous streamer, however, we now utilize optogenetic C2C12 myoblasts.
After fabrication, we use confocal fluorescence microscopy to confirm the expression of ChR2 after differentiation.
Indeed, we observe the persistent expression of ChR2 within the tissue (\cref{fig:3}a) and, therefore, expect the muscle ring to respond to an external light stimulation of wavelength 470~nm. 
An optical fiber is fixed on the microscope stage and is used to uniformly illuminate the muscle ring within the enclosure. 
The fiber delivers 10~ms pulses from the connected laser diode source at driving frequencies of 2, 3, and 4~Hz. 
These driving frequencies are less than, equal to, and greater than the autonomous biostreamer actuation frequency ($\sim 3.1$ Hz).


Upon light stimulation, the representative muscle ring of \cref{fig:3} is observed to contract robustly for all three driving frequencies, as quantified by tracking its edge displacement \cref{fig:3}b.
At the lowest driving frequency (2~Hz), the muscle actuation response is characterized by rapid contractions ($\sim 0.1$~s) followed by slower relaxations ($\sim 0.4$~s) to its original configuration (\cref{fig:3}b, top).
At higher driving frequencies (3 and 4~Hz), the muscle responds to the stimulus consistently, with contractions over $\sim 0.1$ s (\cref{fig:3}b, middle/bottom).
However, the relaxation time ($\leq 0.2$~s) is no longer sufficient for returning to the original configuration and smaller edge displacement amplitudes are realized. 
For the driving frequencies of 2, 3, and 4 Hz, the average peak-to-peak edge displacements are indeed measured to be $81 \ \mu$m, $59 \ \mu$m, and $61 \ \mu$m, respectively. 




Finally, using the same spectral analysis approach of \cref{fig:2}, we characterize the oscillatory and rectified flow fields generated by the optically driven biostreamer, both in simulations and experiments.
For all driving frequencies, we find that the experimental flow (black) responds with hallmarks similar to the spontaneous case, as the dominant mode shifts to align with the optical driving frequency (\cref{fig:3}c), underscoring the system controllability.
Further, and importantly, the zero-frequency peak, which corresponds to the rectified streaming flow, is observed for all three driving frequencies. 
In good agreement with theory and simulations (blue), the magnitude of the zero-frequency peak is an order of magnitude smaller than the dominant mode. 
Moreover, the presence of the cascading peaks at multiples of the driving frequency are evident for all three cases. 
Specifically, for a driving frequency of 2 Hz, peaks of decaying magnitude are observed at 4, 6, 8, and 10 Hz. 
Similarly, for driving frequencies of 3 and 4 Hz, peaks of decaying magnitudes are observed at 6 and 9 Hz, and at 8 and 12 Hz, respectively. 
As for the autonomous biostreamer, the magnitude of the cascading peaks is larger in experiments than simulations, for the same reasons previously described.

\section*{Conclusion}
In this work, we present a bio-hybrid setup that allows us to investigate viscous streaming in biologically-powered, untethered, three-dimensional and millimeter-scale systems, approximating conditions experienced by a host of small aquatic creatures, long speculated (but not rigorously, experimentally confirmed) to autonomously rectify surrounding flows.

By combining tissue engineering, micro-particle image velocimetry, numerical simulations and flow analysis, we design and demonstrate a $\sim$3.5 millimeter muscle ring that, once freely suspended in bulk liquid, generates viscous streaming, autonomously or in a remotely controlled fashion, via light stimulation. This experimentally verifies the speculated ability of millimeter-scale organism to produce and sustain streaming, while paving the way for potential applications in microfluidics and robotics. Further, this work sets the stage for exploring novel streaming flows and dynamics through design and actuation strategies afforded by tissue-engineering approaches \cite{filippi2022will}, broadening the scope of bio-hybrid technology to both fundamental and applied fluid mechanics.



\matmethods{

\subsection*{PDMS Mold}
PDMS base, Sylgard 184, and curing agent (Dow Corning) are mixed with 10:1 ratio by weight and degassed in a vacuum desiccator. 
The mixture is then poured into the machine fabricated aluminum negative mold and cured at 60 $^{\circ}$C overnight. 
The PDMS molds are peeled off from the negative structures the next day and sterilized by autoclave. The PDMS mold has an inner diameter of 5~mm, an outer diameter of 12~mm, and an depth of 3~mm.
Prior to seeding the cell-gel solution, PDMS molds are treated overnight with 1\% (w/v) pluronic F-127 (Sigma) dissolved in phosphate-buffered saline to reduce adhesion of cells and proteins.

\subsection*{Cell culture and muscle ring formation}
C2C12 murine myoblasts and those that are transfected with pLenti2-EF1$\alpha$-ChR2[H134R]-tdTomato-WPRE plasmid to express Channelrhodopsin-2 (ChR2) are seeded in cell culture flasks with growth medium consisted of 10\% fetal bovine serum (Gibco), 1\% L-glutamine (Gibco) and 1\% penicillin-streptomycin (Lonza) in Dulbecco's modified eagle medium (Gibco) and grown to $70-80\%$ confluency. 
For muscle tissue formation, cells are collected and mixed with Matrigel basement membrane matrix and type-I collagen diluted by growth medium. 
The cell-gel mixture is then injected into the PDMS mold and allowed to polymerize for 1 hour at 37$^{\circ}$C, then the sample is inundated in growth medium and incubated for 2 days. 
After 2 days, they are transferred onto hydrogel tubes immersed in differentiation medium, which replaces the fetal bovine serum with horse serum (Gibco) in the growth medium, to induce the formation of myotubes.
The muscle rings are kept on the tubes for 10 days, with the differentiation medium refreshed every 2 days, before being used for the experiment.

\subsection*{Experiment setup}
The aluminum enclosure for the experiment is machined and glued to a microscope glass slide using Kwik-Sil silicone adhesive (World Precision Instrument). 
The box also serves to shield electric field, minimizing tracking particles motion caused by their slight negative charge. 
Before experiment, one muscle ring is transferred from the hydrogel tube to a thin aluminum rod of 1~mm diameter, which is then transferred to the middle of the aluminum enclosure filled with culture media mixed with tracking particles. 
The aluminum enclosure apparatus is then secured on the microscope stage. 
A cover slip is added on top of the box to reduce convection, and the particles are allowed to settle for five minutes before recording. 
Recording is done using a digital complementary metal–oxide semiconductor camera (Hamamatsu) with $2\times$ objective focused to center plane of muscle based on sharpness of the muscle ring edge. 
Each recording is 20 seconds long with 50~Hz capturing frequency, 20~ms exposure time and $2 \times 2$ binning. 
A control experiment, which only have culture media, is conducted with same temperature and carbon-dioxide control as streaming experiment to verify the particles motion without muscle ring is insignificant. 
Optical stimulation is carried out by an optical fiber of diameter 2.6~mm connected to a laser diode (Doric), fixed on the microscope stage next to the aluminum enclosure, producing light pulses at 2, 3 and 4~Hz. 
Each light pulse is 10~ms long with intensity 1.6~mW/mm$^2$. 

\subsection*{Simulation method and numerical implementation}
We briefly recap the governing equations and numerical solution technique. 
We consider incompressible viscous flows in an unbounded domain $\Sigma$. 
In this fluid domain, immersed solid bodies perform simple harmonic oscillations. 
The bodies are density matched and have support $\Omega$ and boundary $\partial\Omega$, respectively. 
The flow can then be described using the incompressible Navier-Stokes equations \cref{eq:1}.

\begin{equation}\label{eq:1}
\nabla \cdot \boldsymbol{u}=0; \quad\frac{\partial u}{\partial t}+(\boldsymbol{u} \cdot \boldsymbol{\nabla}) \boldsymbol{u}=-\frac{\boldsymbol{\nabla} P}{\rho}+\nu \nabla^{2} u, \quad x \in \Sigma \backslash \Omega
\end{equation}

where $\rho$, $P$, $\boldsymbol{u}$ and $\nu$ are the fluid density, pressure, velocity and kinematic viscosity, respectively. 
The dynamics of the fluid-solid system is coupled via the no-slip boundary condition $\boldsymbol{u}=\boldsymbol{u}_{s}$, where $\boldsymbol{u}_{s}$ is the solid body velocity.
The muscle ring contraction is prescribed via a time-varying core radius of a torus $R(t) = R_0 + A\sin(\omega t)$, where $R_0$ is the core radius in rest configuration, and $A$ and $\omega$ are the contraction amplitude and frequency, respectively.
In order to conserve volume, we vary the tube radius of the torus accordingly through $r(t) = r_0\sqrt{R_0 / R(t)}$, where $r_0$ is the tube radius in rest configuration.
Hence, the overall muscle contraction in our simulation is described as a torus with body velocity $dR / dt = A \omega \cos(\omega t)$ accounting for time-varying core radius due to muscle contraction, and body deformation velocity $dr / dt = -0.5 A \omega r_0 R(t)^{-1.5} \sqrt{R_0} \cos(\omega t)$ accounting for tube radius adjustment to conserve volume.
The 3D system of equations is then solved using a velocity-vorticity formulation with a combination of remeshed vortex methods and Brinkmann penalization \cite{gazzola2011simulations}, in the axisymmetric coordinate system. This method has been validated across a range of flow–structure interaction problems, from flow past bluff bodies to biological swimming \cite{gazzola2011simulations,gazzola2012flow,gazzola2012c,gazzola2014reinforcement,gazzola2016learning,bhosale2021remeshed,bhosale2022soft}. Recently, the accuracy of this method has been demonstrated in rectified flow contexts as well, capturing steady streaming responses from arbitrary rigid shapes in 2D and 3D \cite{parthasarathy2019streaming,bhosale_parthasarathy_gazzola_2020,chan2021three,bhosale2021multi}. 

\subsection*{Particle image velocimetry}
Velocity field measurement of streaming is carried out by micro-particle image velocimetry ($\mu$PIV). 
The culture medium is seeded with polystyrene spheroids of diameter 6-8~$\mu$m and specific gravity 1.05. 
The aluminum enclosure minimizes particles motion caused by their slight negative charge. 
A control experiment, in which only the muscle ring is absent while all other conditions are maintained, is first conducted to verify the particle motion without muscle ring is insignificant. 
After the muscle ring is transferred into the system, the microscope is focused to center plane of muscle ring based on sharpness of the its edges. 
1000 images is then collected at a frequency of 50~Hz at 1~MP ($1024 \times 1024$ pixels) resolution. 
The images are then paired with 5 frames step size and then integrated with spatial cross correlation method \cite{hart1998elimination,keane1992theory}. 
The final integration window resulted in $10 \times 10$ pixels with 50\% overlap and a final grid spacing of $\Delta X = \Delta Y = 32.5 ~\mu$m.

\subsection*{Proper orthogonal decomposition}
Dominant coherent structures are obtained with snapshot proper orthogonal decomposition (POD) following Sirovich \cite{sirovich1987turbulence}. 
Velocity fluctuations $\boldsymbol{u}^{\prime}(\boldsymbol{x}, t)$ are decomposed into a deterministic spatially-correlated part $\phi^n(\boldsymbol{x})$ and time-dependent coefficients $a^n(t)$ according to \cref{eq:2}
\begin{equation}\label{eq:2}
\boldsymbol{u}^{\prime}(\boldsymbol{x}, t)=\sum_{n=1}^{N} a^{n}(t) \boldsymbol{\phi}^{n}(\boldsymbol{x})
\end{equation}
where $N\approx 1000$ is the number of snapshot. 
According to corresponding ratio of the eigenvalues to summation of $N$ eigenvalues, i.e. $E_{n}=\lambda_{n} / \sum_{m=1}^{N} \lambda_{m}$, each mode is sorted by its contribution to the turbulence kinetic energy. 

\subsection*{Spectral analysis}
For the muscle ring, the displacement of the top and bottom edges is tracked using the image analysis software Tracker(physlets.org/tracker). An FFT is then performed on the displacement data to produce the spectrum of muscle actuation. For the experimental flow, one point within the recirculation region is selected and FFT is performed on its time-dependent POD coefficients. 

\subsection*{Muscle ring staining}
The muscle ring is rinsed with PBS and fixed in 4\%v/v of paraformaldehyde for 30 minutes. It is then washed three times for 5 minutes to permeabilize the tissue sample and incubated with 0.2\%v/v Triton X-100 (Sigma) diluted in PBS for 15 min. The muscle ring is then blocked and stored in Image-iT FX Signal Enhancer blocking solution (Invitrogen) at 4$^\circ$C overnight.
The primary antibody, mouse antimyosin heavy chain (MF-20) is used to stain for myosin heavy chain with a 1:400 dilution ratio. The sample is incubated overnight at 4$^\circ$C and then washed three times for 5 minutes before staining with secondary antibodies on the next day. The secondary antibody, AlexaFluor-488 anti-mouse (Invitrogen) is used to stain the MF-20 antibody overnight. The sample is incubated with DAPI (Invitrogen) for 15 minutes to stain nuclei at 4$^\circ$C. After washing with PBS three times, the LSM 700 is used for the confocal fluorescent imaging.

}

\showmatmethods{} 

\acknow{The authors thank S. Hilgenfeldt for helpful discussions over the course of this work. The authors acknowledge support by the National Science Foundation under NSF CAREER Award \#CBET–1846752 (MG) and the NSF Expedition ‘Mind in Vitro’ Award \#IIS–2123781 (MG, TS).}

\showacknow{} 

\bibliography{pnas-sample}

\end{document}